\documentclass{article}

\usepackage{arxiv}

\usepackage[utf8]{inputenc} 
\usepackage[T1]{fontenc}    
\usepackage{hyperref}       
\usepackage{url}            
\usepackage{booktabs}       
\usepackage{amsfonts}       
\usepackage{nicefrac}       
\usepackage{microtype}      
\usepackage{lipsum}
\usepackage{array}
\usepackage{graphicx}
\graphicspath{ {./images/} }

\usepackage{lmodern}
\usepackage{textcomp}
\usepackage{spverbatim}

\usepackage{amsmath, amssymb, amsfonts}

\usepackage[table]{xcolor} 
\definecolor{PineGreenDarkest}{HTML}{4C9E8F}
\definecolor{PineGreenDark}{HTML}{70B3A1}
\definecolor{PineGreenMedium}{HTML}{94C8B4}
\definecolor{PineGreenLight}{HTML}{B8DDC6}
\definecolor{PineGreenLighter}{HTML}{D1EED9}
\definecolor{RedLight}{HTML}{F8D7DA}
\definecolor{darkgreen}{RGB}{0,100,0} 

\usepackage{graphicx}
\usepackage{float}

\usepackage{subcaption}

\usepackage{multirow}
\usepackage{tabularx}
\usepackage{tabularray}
\usepackage{colortbl}
\usepackage{makecell}

\usepackage{algorithmic}

\usepackage{fancyvrb}
\usepackage[most]{tcolorbox}
\usepackage{minted}   
\usepackage{listings} 

\lstdefinestyle{mypython}{
    language=Python,
    basicstyle=\ttfamily\small,
    keywordstyle=\color{blue}\bfseries,
    stringstyle=\color{darkgreen},
    commentstyle=\color{gray}\itshape,
    morekeywords={BaseModel, Field, List},
    columns=flexible,
    keepspaces=true,
    breaklines=true,
    showstringspaces=false,
    frame=single,
    rulecolor=\color{black},
    stepnumber=1,
    numbersep=5pt,
    xleftmargin=0.05\columnwidth,
    xrightmargin=0.05\columnwidth,
}
\lstdefinestyle{promptblock}{
  basicstyle=\ttfamily\small,
  frame=single,
  breaklines=true,
  breakatwhitespace=true,
  breakautoindent=false,
  breakindent=0pt,
  prebreak=\mbox{},
  postbreak=\mbox{},
  columns=fullflexible,
  keepspaces=true,
}

\title{LM4Opt-RA: A Multi-Candidate LLM Framework with Structured Ranking for Automating Network Resource Allocation}

\author{
 Tasnim Ahmed\\
  School of Computing\\
  Queen's University\\
  Kingston, Ontario, Canada K7L 2N8\\
  \texttt{tasnim.ahmed@queensu.ca}\\
   \And
 Siana Rizwan\\
  School of Computing\\
  Queen's University\\
  Kingston, Ontario, Canada K7L 2N8\\
  \texttt{siana.rizwan@queensu.ca}\\
  \And
 Naveed Ejaz\\
  School of Computing\\
  Queen's University\\
  Kingston, Ontario, Canada K7L 2N8\\
  \texttt{ht57@queensu.ca}\\
  \And
  Salimur Choudhury\\
  School of Computing\\
  Queen's University\\
  Kingston, Ontario, Canada K7L 2N8\\
  \texttt{s.choudhury@queensu.ca}\\
}

\begin{document}
\maketitle
\begin{abstract}
Building on advancements in Large Language Models (LLMs), we can tackle complex analytical and mathematical reasoning tasks requiring nuanced contextual understanding. A prime example of such complex tasks is modelling resource allocation optimization in networks, which extends beyond translating natural language inputs into mathematical equations or Linear Programming (LP), Integer Linear Programming (ILP), and Mixed-Integer Linear Programming (MILP) models. However, existing benchmarks and datasets cannot address the complexities of such problems with dynamic environments, interdependent variables, and heterogeneous constraints.
To address this gap, we introduce NL4RA, a curated dataset comprising 50 resource allocation optimization problems formulated as LP, ILP, and MILP. We then evaluate the performance of well-known open-source LLMs with varying parameter counts. To enhance existing LLM based methods, we introduce LM4Opt RA, a multi candidate framework that applies diverse prompting strategies such as direct, few shot, and chain of thought, combined with a structured ranking mechanism to improve accuracy.
We identified discrepancies between human judgments and automated scoring such as ROUGE, BLEU, or BERT scores. However, human evaluation is time-consuming and requires specialized expertise, making it impractical for a fully automated end-to-end framework. To quantify the difference between LLM-generated responses and ground truth, we introduce LLM-Assisted Mathematical Evaluation (LAME), an automated metric designed for mathematical formulations. Using LM4Opt-RA, Llama-3.1-70B achieved a LAME score of 0.8007, outperforming other models by a significant margin, followed closely by Llama-3.1-8B. While baseline LLMs demonstrate considerable promise, they still lag behind human expertise; our proposed method surpasses these baselines regarding LAME and other metrics.
\end{abstract}


\section{Introduction}
Networks are evolving to support a wide range of heterogeneous applications in increasingly diverse and dynamic environments. These modern networks accommodate IoT devices, vehicular communications, mobile edge computing, automated industries, remote telemedicine, and smart cities \cite{damsgaard2023approximate, hasan2024federated}. Consequently, the demand for faster, more reliable, and low-latency connections has intensified, as these services require varying levels of Quality of Service (QoS) \cite{wang2023road}. To achieve maximum performance and provide a good user experience in networks, their resources must be dynamically and efficiently allocated to meet the continuously changing demands regarding power control, bandwidth allocation, deployment strategies, association allocation, etc.\cite{salahdine20235g, sarah2023resource}. The performance of a network largely depends on how its resources are allocated. Linear Programming (LP), Integer Linear Programming (ILP), and Mixed-Integer Linear Programming (MILP) are commonly used to formulate resource allocation problems in networks \cite{alhashimi2023survey}. LP addresses problems with continuous variables, ILP handles problems with integer variables, and MILP tackles problems involving both continuous and integer variables \cite{taha2013operations}.

Formulating a network resource allocation problem into LP, ILP, or MILP requires significant mathematical expertise and in-depth knowledge of the problem domain, and it is often time-consuming \cite{chakraa2023optimization}. With the growing popularity of Large Language Models (LLMs), some researchers have investigated methods to automate the formulation of optimization problems using LLMs. Some of the recent works suggest that LLMs can formulate and solve linear programming problems adequately using natural language description \cite{li2023synthesizing, ramamonjison2023nl4opt, ahmed2024lm4optunveilingpotentiallarge, tsouros2023holy}. This early success of LLMs indicates their potential to simplify the process of problem formulation and solution in operations research, including network resource allocation, which falls under the broader category of similar mathematical optimization problems. By automating the generation and solving of mathematical formulation, LLM-based frameworks can effectively reduce the need for extensive mathematical expertise, allowing professionals or stakeholders to focus on listing resources and constraints. This approach lowers the resource allocation cost and improves the efficiency and accuracy of translating complex real-world problems into solvable mathematical formats, advancing quicker solution development and innovation.

Existing benchmarks for optimization modeling, such as NL4Opt, MAMO, and IndustryOR, focus predominantly on generalized optimization tasks, often constrained to LP or elementary MILP problems. These datasets fail to encompass the heterogeneity and real-time adaptation requirements inherent in modern network resource allocation. Furthermore, current LLM-based frameworks exhibit limitations in generating solver-specific code and handling unstructured problem descriptions. Consequently, there remains a critical gap in addressing network-specific optimization challenges with the precision and adaptability required for practical deployment.

\begin{figure*}[htbp]
    \centering
    \includegraphics[width=\textwidth]{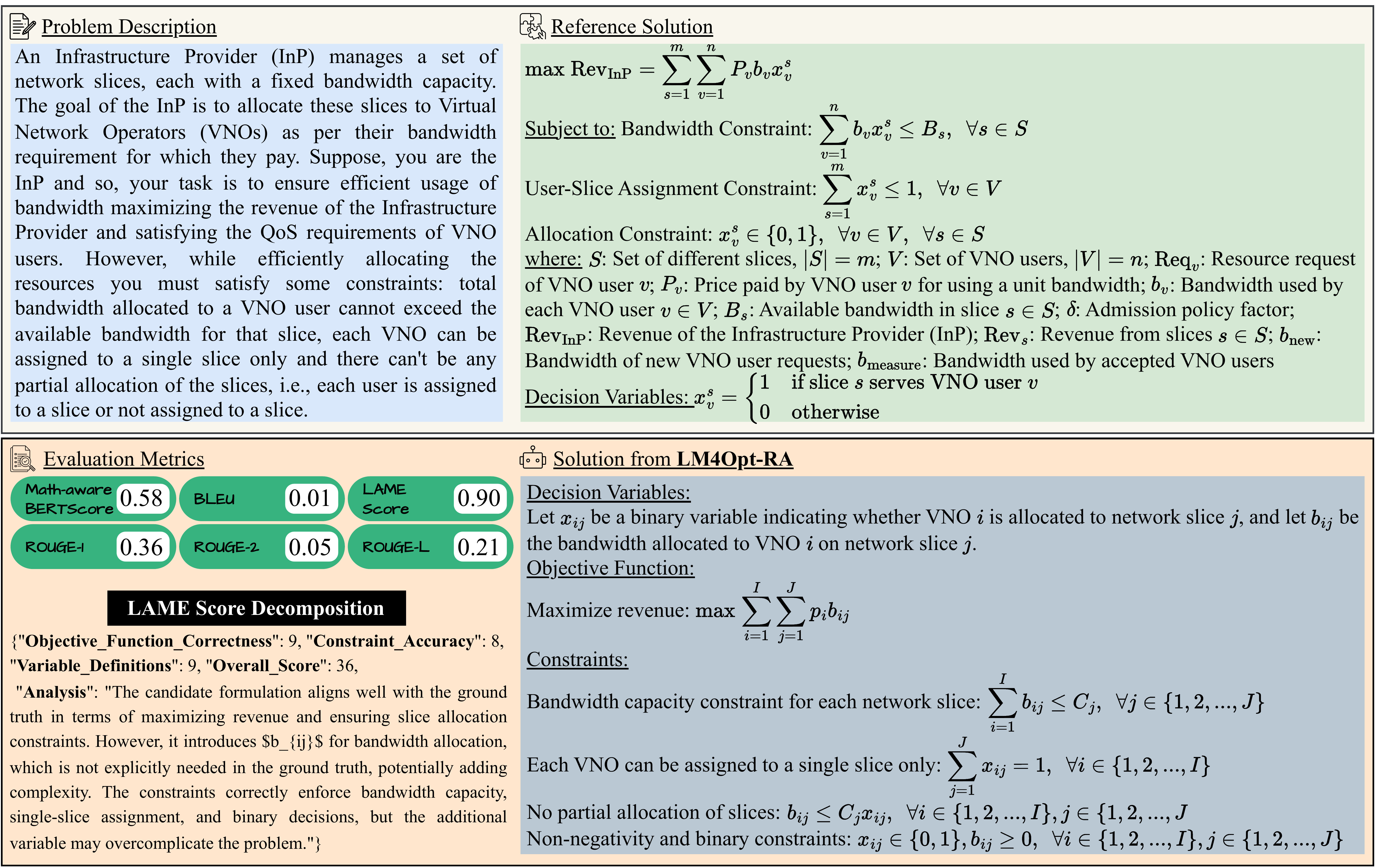}
    \caption{Translation of a natural language resource allocation problem into a mathematical formulation using LM4Opt-RA, highlighting the distinction between traditional metrics and the proposed LAME Score}
    \label{fig:overview}
\end{figure*}

To address these challenges, we introduce NL4RA, a curated dataset comprising $50$ peer-reviewed research-based optimization problems designed for network resource allocation. The dataset spans diverse domains, including 5G/6G technologies, edge computing, and software-defined networking (SDN), offering mathematical formulations for LP, ILP, and MILP problems. Building on this dataset, we propose LM4Opt-RA, a multi-candidate LLM framework designed to improve the translation of natural language descriptions into mathematical formulations. The framework adopts diverse prompting strategies---direct, few-shot, and chain-of-thought---to generate multiple candidate solutions, followed by a structured ranking mechanism to identify the most accurate formulation. Our evaluation implements both traditional and advanced metrics, including BLEU, ROUGE, Math-aware BERTScore, and LLM-Assisted Mathematical Evaluation (LAME) score. These metrics evaluate the semantic and structural correctness of mathematical formulations. We evaluate the proposed framework with two open-source LLMs at different parameter scales and temperature settings. In our empirical evaluation, the proposed framework exhibited superior performance compared to the baseline LLMs. For instance, at a temperature setting of $0.0$, the proposed framework enhanced LAME-5 scores for Llama-3.1-70B to $0.8007$, substantially improving over the baseline ($0.7492$). Figure~\ref{fig:overview} illustrates an example of translating a natural language description of a resource allocation problem into a mathematical formulation using our proposed framework, LM4Opt-RA. Additionally, we emphasize the distinction between traditional automated evaluation metrics and our novel LAME Score, which evaluates the framework-generated solution against the ground truth.

The contributions of this study can be summarized as:
\begin{enumerate}
    \item We introduce NL4RA, a curated dataset of 50 real-world network resource allocation problems (LP, ILP, and MILP) capturing the heterogeneity and complexity overlooked in previous benchmarks.
    \item A multi-candidate LLM approach, LM4Opt-RA, is proposed that integrates direct, few-shot, and chain-of-thought prompts with a structured ranking mechanism for more accurate mathematical formulations.
    \item We develop a math-aware evaluation metric, LAME, which evaluates correctness and completeness more effectively than standard text-overlap measures.
    \item An empirical study is conducted to demonstrate that our approach outperforms baseline LLMs on the new dataset, showing improved formulation accuracy and alignment with human judgments.
\end{enumerate}

The remainder of this paper is organized as follows: \textbf{Related Work} reviews prior efforts on using LLMs for optimization modeling and identifies gaps in network resource allocation. \textbf{System Model} outlines the system architecture and its components, emphasizing the integration of LLM-based resource allocation frameworks. \textbf{NL4RA Dataset} details the dataset’s design methodology, problem categorization, and features tailored to network optimization challenges. In \textbf{Mathematical Formulations with a Multi-Candidate LLM Framework}, we present LM4Opt-RA, elaborating on multi-candidate generation and structured ranking. \textbf{Evaluation Metrics} discusses the metrics used, including the novel LAME score, and highlights their alignment with human judgments. \textbf{Findings and Analysis} provides a comprehensive evaluation of experimental results, comparing LM4Opt-RA with baseline models and discussing limitations. Finally, \textbf{Conclusion} summarizes key findings, contributions, and directions for future research in LLM-driven optimization for network resource allocation.

\section{Related Work}
\label{sec:related_work}
Recently, several approaches have been proposed to bridge the gap between natural language descriptions and mathematical optimization problems. 
Ramamonjison et al. \cite{ramamonjison2023nl4opt} introduced the NL4Opt Competition focused on identifying linked entities within optimization problem descriptions and generating the corresponding mathematical formulations. Extending one of the subtasks of the competition, Dakle et al. \cite{dakle2023ner4opt} proposed a hybrid approach that combines lexical and semantic models, enhanced by feature engineering, classical NER techniques, and data augmentation techniques, to enhance performance in recognizing optimization-related entities. However, both works primarily addressed LP problems, and the curated dataset comprised elementary optimization problem samples that do not adequately represent real-world scenarios. Additionally, the competition did not involve generating solver-specific code. Building on the contributions of NL4Opt, Ahmed et al. \cite{ahmed2024lm4optunveilingpotentiallarge} evaluated various LLMs for converting linguistic descriptions to mathematical problem formulation, introducing the LM4OPT framework. Their study demonstrated that GPT-4 outperformed other models, achieving $63.30\%$ accuracy without additional named entity information. Fine-tuning smaller models with LM4OPT was shown to narrow the performance gap between smaller and larger models. However, their findings were based solely on the NL4Opt dataset, limiting generalizability to more complex, real-world problems.

Li et al. \cite{li2023synthesizing} extended the NL4Opt framework by incorporating MILP problems and handling logic constraints. Their approach involved expanding the NL4Opt dataset to include binary variables and various logic constraints and proposing a three-stage framework using LLMs for variable identification, classification, and generation. Despite these advancements, the dataset remained limited in diversity regarding contexts, constraint types, and optimization problem types and did not cover solver-specific code generation.

AhmadiTeshnizi et al. \cite{ahmaditeshnizi2023optimus} presented OptiMUS, an LLM-based agent for generating mathematical formulations and solver code, with automated testing and debugging features, and data augmentation through problem rephrasing. They introduced the NLP4LP dataset, which includes LP and MILP problems from textbooks and lecture notes. While providing solutions and code for optimality checks, the NLP4LP dataset remains relatively small and may not fully capture the complexity of real-world problems. Moreover, OptiMUS relied on structured natural language representations, limiting its applicability to unstructured problem descriptions. Xiao et al. \cite{xiao2023chain} introduced the Chain-of-Experts framework, employing multiple LLM-based agents for model construction, programming, and code review, coordinated by a ``Conductor" with a reflection mechanism for error correction. They also introduced ComplexOR, a new benchmark with 37 diverse problems, including expert annotations and model formulations. However, using multiple LLM agents increases computational costs, and the ComplexOR dataset samples may not fully represent real-world problem diversity. Huang et al. \cite{huang2024mamo} proposed MAMO, a benchmark evaluating the mathematical modeling capabilities of LLMs by focusing on the underlying modeling process rather than just assessing the correctness of final responses. The dataset includes 346 ordinary differential equations problems, 652 easy, and 211 complex LP problems. However, MAMO's focus on LP problems limits its applicability to a broader range of optimization problems. Yang et al. \cite{yang2024benchmarking} introduced E-Opt, a benchmark requiring LLMs to generate Python code that utilizes optimization solvers. It covers a range of optimization problems, including LP, MILP, and quadratic programming, with varying difficulty levels. The study explored different prompting strategies and demonstrated the benefits of fine-tuning LLMs on domain-specific datasets. Nonetheless, the E-Opt benchmark remains relatively small in scale.

Tang et al. \cite{tang2024orlm} proposed a semi-automated process for creating synthetic data to train open-source LLMs for optimization modeling tasks. They introduced IndustryOR, the first industrial benchmark for evaluating LLMs on real-world optimization problems. Their results showed that fine-tuned open-source LLMs achieved accuracy rates of $85.7\%$ on NL4Opt, $82.3\%$ on MAMO (Easy LP), and $37.4\%$ on MAMO (Complex LP). However, the data generation process may introduce biases, and the study focused on 7b-size LLMs, leaving other larger models unexplored. Furthermore, Mostajabdaveh et al. \cite{mostajabdaveh2024evaluatingllmreasoningoperations} introduced Operations Research Question Answering (ORQA), a benchmark aimed at evaluating the generalization capabilities of LLMs in the specialized domain of operations research (OR). ORQA drafted by OR experts, presents complex optimization problems that require multi-step reasoning. They evaluated various open-source LLMs, including LLaMA 3.1, DeepSeek, and Mixtral, revealing modest performance, with LLaMA 3.1-405B-Instruct achieving the highest accuracy of $0.772$ compared to human expert accuracy of $0.93$.  Nonetheless, a more detailed evaluation of diverse real-world scenarios is needed to evaluate LLMs' capabilities as both works suggest.

\begin{table*}[t]
\centering
\caption{Datasets Containing Optimization Problems for LLM Modeling}
\label{tab:Datasets}
\renewcommand{\arraystretch}{1.2}
\begin{tabular}{|>{\arraybackslash}m{2.7 cm}|>{\centering\arraybackslash}m{2.3cm}|>{\centering\arraybackslash}m{0.6cm}|>{\centering\arraybackslash}m{0.6cm}|>{\centering\arraybackslash}m{0.9cm}|m{7.5cm}|}
\hline
\textbf{Dataset Name} & \textbf{\# Problems} & \textbf{LP} & \textbf{ILP} & \textbf{MILP} & \textbf{Optimization Problem Types} \\
\hline
NL4Opt  \cite{ramamonjison2023nl4opt} & 1101 & \checkmark & \texttimes & \texttimes & Sales, Advertising, Investment, etc. \\
\hline
Mamo Easy \cite{huang2024mamo} & 652 & \texttimes & \texttimes & \checkmark & High school-level textbook problems \\
\hline
Mamo Complex \cite{huang2024mamo} & 211 & \checkmark & \texttimes & \texttimes & Undergrad level textbook problems \\
\hline
IndustryOR \cite{tang2024orlm} & 100 & \checkmark & \checkmark & \checkmark & General Industry problems \\
\hline
NLP4LP \cite{ahmaditeshnizi2023optimus} & 52 & \checkmark & \texttimes & \checkmark & Text Books \\
\hline
ComplexOR \cite{xiao2023chain} & 37 & \texttimes & \texttimes & \checkmark & Research Papers, Textbooks, Industry \\
\hline
\textbf{Proposed} & 50 & \checkmark & \checkmark & \checkmark & \textbf{Network Resource Allocation Optimization} \\
\hline
\end{tabular}
\end{table*}

As seen in Table \ref{tab:Datasets}, the existing datasets focused mainly on using LLMs for general optimization tasks, with limited or no attention to the specific challenges of network resource allocation. Most datasets have textbook optimization problems and are relatively easier to understand. The network resource optimization problem is a specialized niche with additional complexity. The optimization problems in modern network resource allocation problems are usually more complex than traditional optimization problems contained in the existing datasets. Unlike standard textbook optimization problems, network resource allocation problems are characterized by a dynamic probabilistic ecosystem with multiple interdependent variables, real-time adaptation requirements, and exponential computational complexity. There are usually heterogeneous constraints such as bandwidth limitations, latency requirements, energy efficiency, security protocols, and quality of service metrics. Moreover, network resource allocation must continuously balance multiple conflicting objectives—maximizing throughput while minimizing latency, ensuring security without compromising performance, and maintaining energy efficiency—all within a constantly evolving technical environment. Our study addresses this gap by providing a dataset specifically for network resource allocation and introducing a natural language interface framework tailored to solve network resource allocation problems. 

\section{System Model}
\label{sec:System_Model}
\begin{figure}[ht]
    \centering
    \includegraphics[width=\columnwidth]{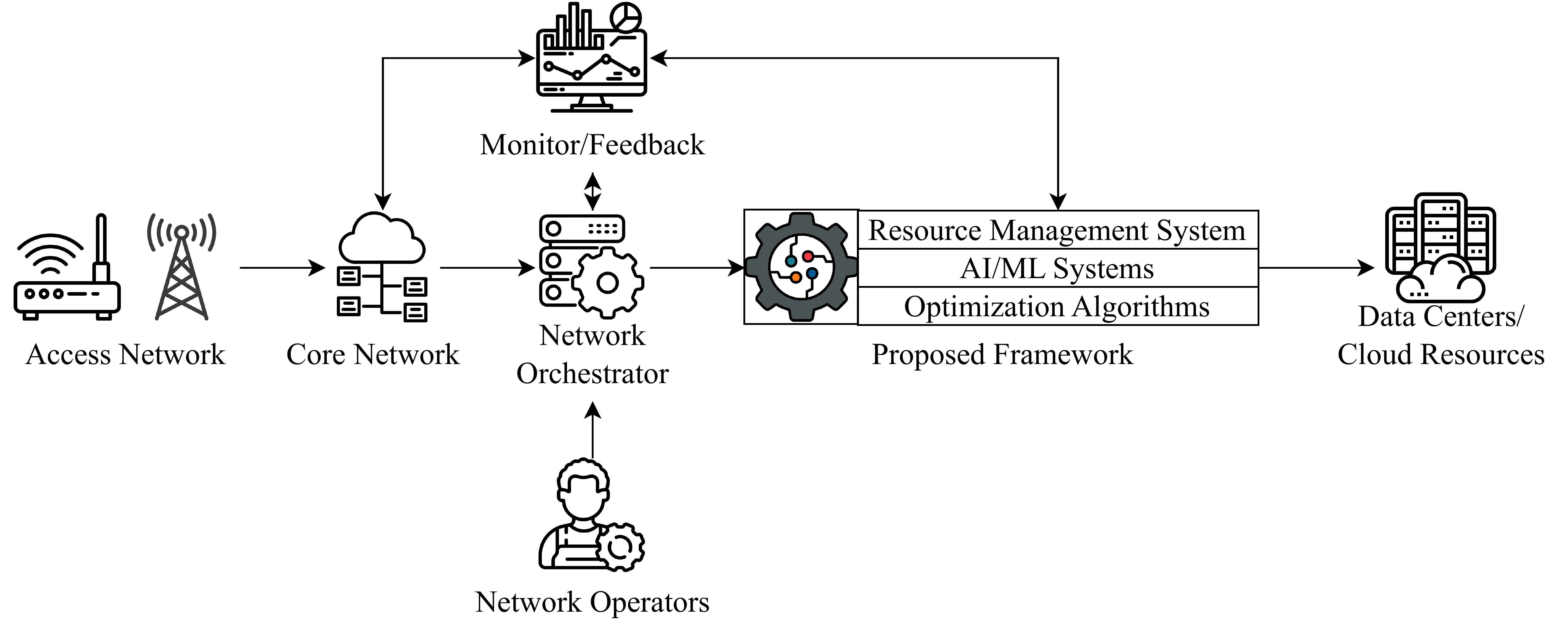}
    \caption{\textbf{High-level System Model.} The system model integrates an AI-driven resource allocation framework that combines resource management systems, AI/ML, and optimization algorithm solvers to manage network resources dynamically. Data from the access and core networks is processed by the LLM-based proposed resource allocation framework, which uses real-time monitoring and feedback to optimize resource allocation. Network operators oversee the process, implementing final decisions in data centers or Cloud resources.}
    \label{fig:system}
\end{figure}

A high-level system model representing a use case of the proposed LM4Opt-RA framework is depicted in \figureautorefname~\ref{fig:system}. 
The access and core networks are the foundational layers responsible for handling data transmission and routing between user requests and the network infrastructure. These layers provide crucial inputs to the network orchestrator, which oversees the overall management and flow of resources across the system.
Central to this model is the LLM-based resource allocation framework, which is responsible for generating and solving optimization problems and managing resources using AI-based methods.
It is responsible for dynamically adjusting and allocating resources based on real-time inputs, predictive analytics, and optimization techniques. The monitoring/feedback loop continuously gathers performance data from various network components and feeds it back into the optimization process, ensuring the system adapts to changing conditions efficiently. Resource allocation decisions are made iteratively, with the system checking whether the assignments are satisfactory through the monitoring loop. If the allocation is not optimal, expert (network operators) feedback is considered, and adjustments are made before arriving at the final allocation of resources, e.g., hosted in data centers or on the cloud. Network operators oversee the entire system, provide human oversight, and intervene when necessary to adjust policies or parameters in the LM4Opt-RA framework.

\section{NL4RA Dataset}
\label{sec:Proposed_Dataset}
\begin{figure*}[htbp]
    \centering
    \includegraphics[width=\textwidth]{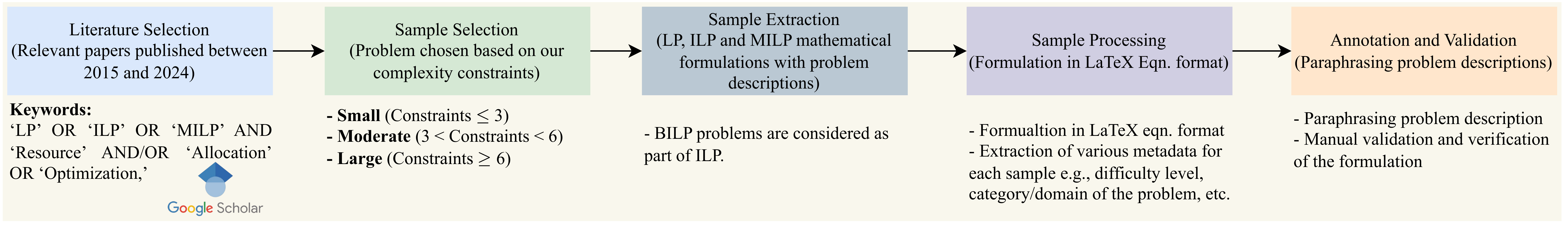}
    \caption{Dataset Preparation Process}
    \label{fig:NL4RA_Dataset_Prep}
\end{figure*}

 The proposed NL4RA dataset is a collection of mathematical models and natural language problem descriptions covering a wide range of cellular networking scenarios. For the dataset, we systematically reviewed the literature, including publications from 2015 to 2024. The works were selected from Google Scholar, emphasizing keywords like `LP' OR `ILP' OR `MILP' AND `Resource'  AND/OR `Allocation' OR `Optimization,'. This resulted in a diverse pool of relevant research works focusing on LP, ILP, and MILP problems as detailed in Table \ref{Tab-SelectedPapers-Problem Formulation Types}. The ILP problems further include a good number of BILP problems as well. The categorization of selected works based on the wide range of networking domains is provided in Appendix A. Each selected paper was extensively analyzed to extract the mathematical formulations and their corresponding problem descriptions. The problem descriptions and formulations are prepared in LATEX format.  The mapping of the selected works with the sample instances extracted from it for NL4RA is also present in Appendix A. The mathematical formulations include a list of variables, constraints, and problem descriptions. The natural language problem descriptions were restructured for clarity and easy interpretability. \figureautorefname~\ref{fig:NL4RA_Dataset_Prep} provides an overview of the overall dataset preparation process.
 
\begin{figure}[t]
    \centering
    \includegraphics[width=0.6\columnwidth]{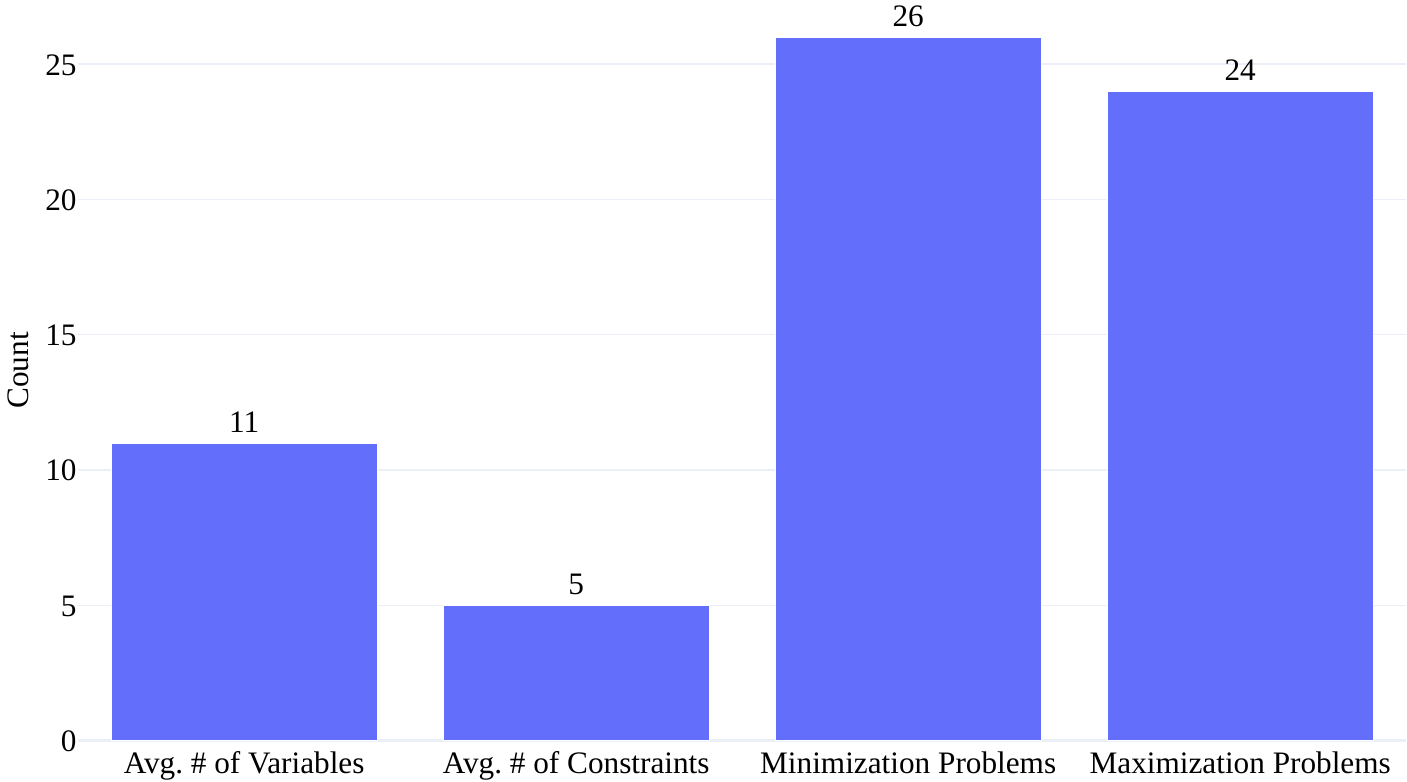}
    \caption{Statistical summary of the NL4RA dataset}
    \label{fig:NL4RA_Overview}
\end{figure}

\begin{figure}[t]
    \centering
    \includegraphics[width=\columnwidth]{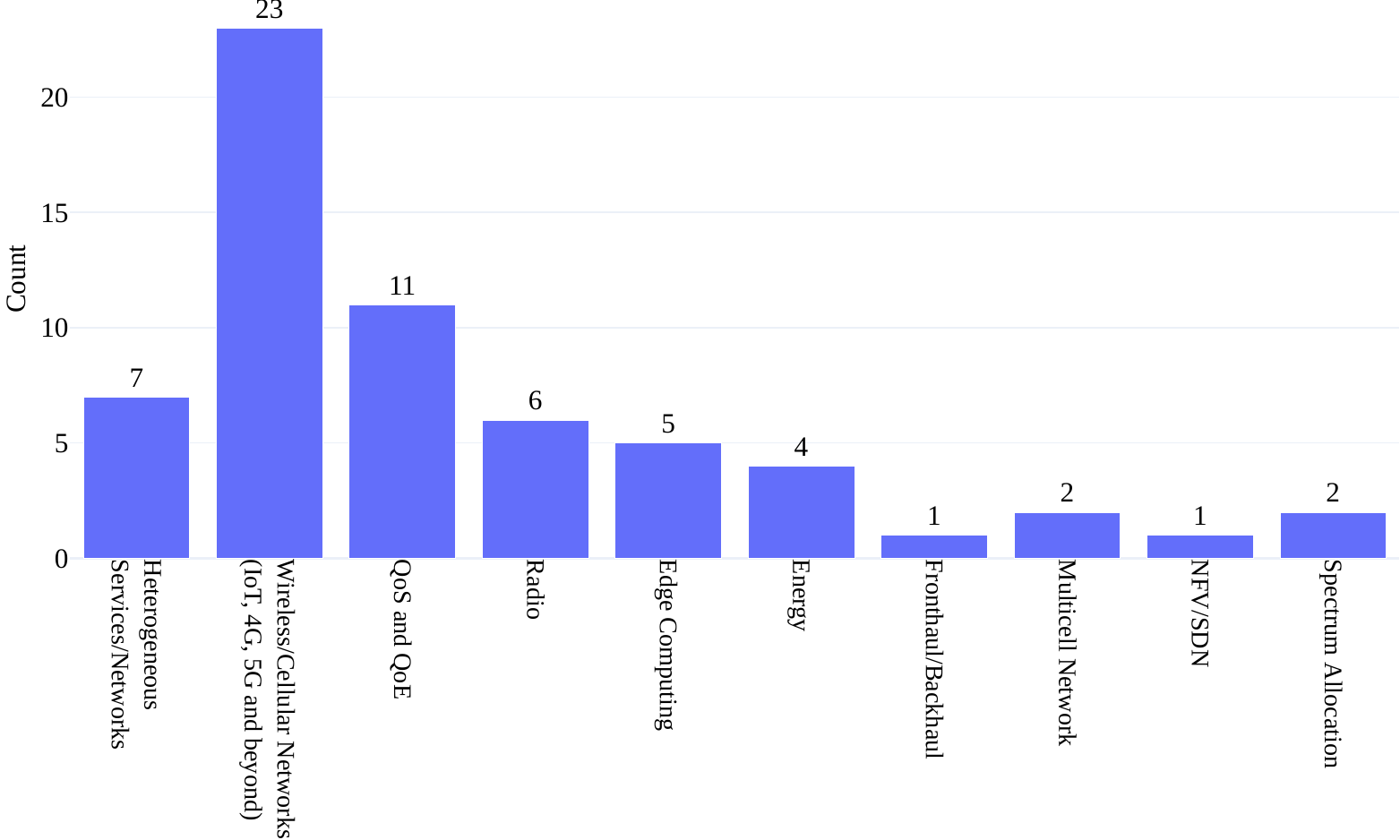}
    \caption{Categorization of dataset samples based on resource allocation types}
    \label{fig:sample_domain}
\end{figure}

NL4RA consists of $50$ unique mathematical models with their problem descriptions. An overview of the attributes of NL4RA is shown in \figureautorefname~\ref{fig:NL4RA_Overview}, indicating that, on average, each problem instance contains 11 variables and 5 constraints, with a total of 26 minimization problems and 24 maximization problems. Furthermore, \figureautorefname~\ref{fig:sample_domain} categorizes the samples of the dataset into distinct resource allocation types. Each category includes samples depicting the diversity of resource management challenges in telecommunications and networking. 

\begin{figure}[t]
    \centering
    \includegraphics[width=0.6\columnwidth]{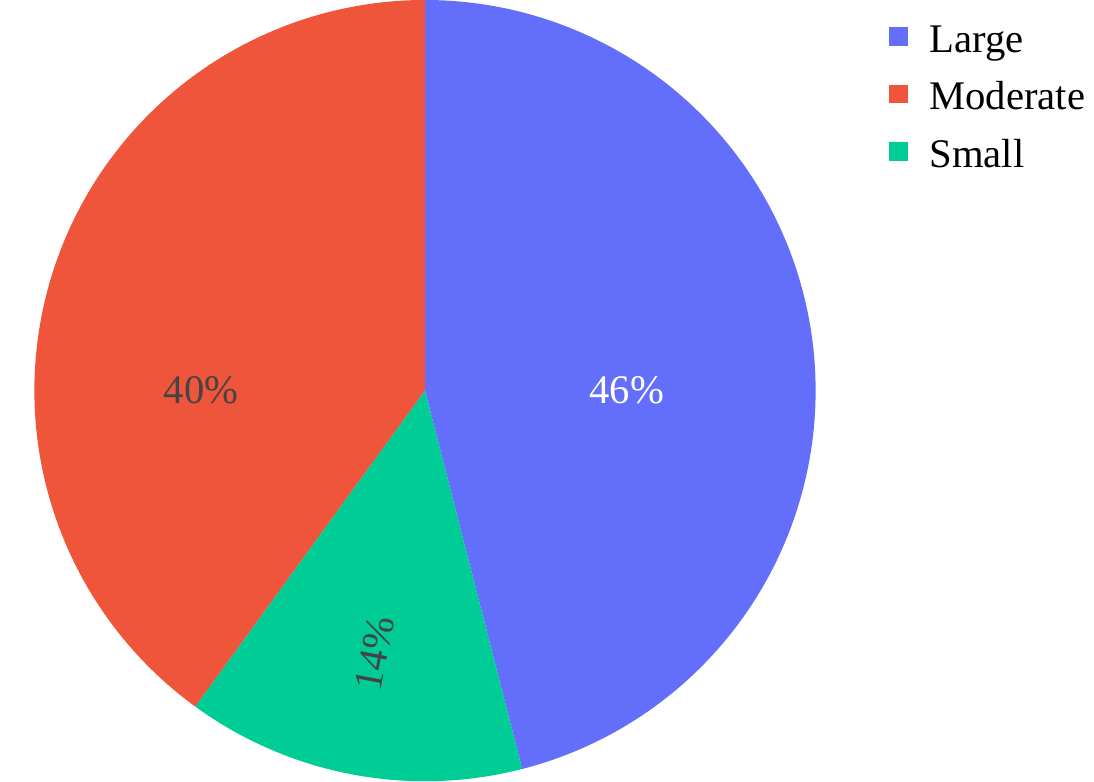}
    \caption{Categorization of samples based problem size}
    \label{fig:problem_difficulty_NL4RA}
\end{figure}

Given the limitations of current LLMs in performing logical and mathematical reasoning \cite{mirzadeh2024gsm}, we imposed restrictions on the complexity of the problem scenarios in NL4RA. We chose problem scenarios with a count of variables not more than 20 and constraints not more than 10. To further restructure the dataset, we have categorized the size of problem instances into Small (constraints not more than 3), Moderate (constraints more than 3 but less than 6), and Large (constraints equal to or more than 6) based on the number of constraints as detailed out in \figureautorefname~\ref{fig:problem_difficulty_NL4RA}. However, there are multiple instances where multiple constraints are grouped under one parent constraint. These multiple constraints were considered individual constraints while the samples were being categorized. 

The creation of NL4RA required extensive analysis and integration of diverse research works into a unified, standardized benchmarking dataset. We expect that NL4RA will offer a foundation for benchmarking LLMs in resource allocation tasks, enabling advancements in networking and AI research domains.

\begin{table}[t]
\centering
\caption{Categorization of Selected Works for NL4RA based on Problem Formulation Types}
\label{Tab-SelectedPapers-Problem Formulation Types}
\begin{tabular}{|p{1.8cm}|p{12cm}|}

\hline
\textbf{Types} & \textbf{References} \\
\hline
LP &  \cite{saghezchi2017energy}, \cite{you2018resource}, \cite{ko2022pdras}, \cite{yang2019two},\cite{8204056}, \cite{alnakhli2024optimizing}\\
\hline
ILP & \cite{hassan2017interference}, \cite{spinelli2022edge}, \cite{ferdosian20225g}, \cite{sharara2022policy}, \cite{zhai2017adaptive}, \cite{javad2021re}, \cite{feng2017adaptive}, \cite{xie2017energy}, \cite{debbabi2022inter}, \cite{emu2021ensemble}, \cite{fayad20235g}, \cite{pan2020multi}, \cite{yang2020towards}, \cite{fendt2018network}, \cite{jia2017efficient}, \cite{9827120}, \cite{almasaeid2023minimum}, \cite{alsheyab2019near}, \cite{spinelli2022migration}, \cite{kim2022modems}, \cite{mharsi2018joint}, \cite{hussain2018system},\cite{hossen2018relax}, \cite{hassan2019online}, \cite{vlachos2017moca}, \cite{9068264}, \cite{de2020optimal}, \cite{salameh2022opportunistic}, \cite{zoha2017leveraging}\\
\hline
MILP &  \cite{de2020optimal}, \cite{hadi2020patient}, \cite{gao2019deep}, \cite{sattar2019optimal}, \cite{yu2019network}, \cite{9929619}, \cite{alam2021resource}, \cite{al2021joint}, \cite{shokrnezhad2018joint}, \cite{you2017load}\\
\hline
\end{tabular}
\end{table}


\section{Mathematical Formulations with a Multi-Candidate LLM Framework}
\label{sec:methodology}
\subsection{Task Definition}
For the task of formulating network resource allocation optimization problem, we use a corpus of network resource allocation problems derived from peer-reviewed research articles and evaluate the capability of LLMs to generate their corresponding mathematical formulations. We focus on complete problem formulations rather than standalone equation generation, as they provide a comprehensive context and better represent real-world applications of resource allocation modeling. The original corpus of problem descriptions is denoted as $P = \{p_1, p_2, \ldots, p_n\}$, where $n = 50$. We obtain mathematical formulations for these problems denoted as $M = \{f(p_1), f(p_2), \ldots, f(p_n)\}$, where $f$ represents LM4Opt-RA framework that transforms natural language problem descriptions into mathematical formulations. Each problem formulation $m \in M$ consists of three essential components, i.e., $m = \{V, C, O\}$, where $V$ represents the set of variable definitions, $C$ denotes the set of constraints, and $O$ represents the objective function. Each component is expressed in standardized mathematical notation following LP, ILP, or MILP conventions. Our secondary objective is to evaluate how accurately the generated formulations $M$ capture the mathematical relationships and optimization goals presented in the original problem descriptions while maintaining mathematical correctness and completeness.

\subsection{LM4Opt-RA}
In contrast to approaches that rely on a single generative step to translate a natural language description into a mathematical formulation, we hypothesize that the quality of these formulations can be substantially improved by incorporating multiple prompts and a ranking mechanism. Instead of forcing an LLM to memorize or guess the best strategy for every type of resource allocation problem, we propose first to generate multiple candidate solutions ---each guided by a different prompting strategy—and then select the most suitable candidate through a subsequent ranking procedure. To this end, we introduce LM4Opt-RA, a multistage framework composed of three main prompts: (i) a direct LLM query, (ii) a few-shot strategy, (iii) a chain-of-thought prompt, and (iv) a final ranking component that evaluates and ranks the candidate formulations. In doing so, LM4Opt-RA ensures that each candidate formulation is informed by a distinct prompting strategy which allows it to capture diverse constraints and contextual nuances.

\subsection{Generating Multiple Candidate Formulations}
Let \(T\) be a textual problem description that describes domain-specific details. We aim to create an LP/ILP/MILP model that aligns with \(T\). Rather than treating this generation process as a single-prompt task, we produce a set of candidate formulations: $F(T) = \{f_{1}(T), f_{2}(T), f_{3}(T)\}$,
where \(f_{1}(T)\), \(f_{2}(T)\), and \(f_{3}(T)\) refer respectively to the formulations generated by the direct, few-shot, and chain-of-thought strategy. The direct prompt query is constructed as a minimal set of instructions focusing on equation generation and the prohibition of extraneous text. It resembles a zero-shot or near-zero-shot scenario, where the LLM is instructed to directly translate \(T\) into a coherent optimization model. In contrast to the direct query, a few-shot prompt demonstrates how a sample resource allocation description (including decision variables, constraints, and an objective function) translates into a final formulation. By providing a structured, validated example, the LLM is more likely to produce a well-structured solution, complete with appropriate indexing, standard notations, and alignment between constraints and the objective function. The third candidate formulation, \(f_{3}(T)\) is derived using a chain-of-thought strategy. Rather than specifying only the final output format, this strategy instructs the LLM to systematically reason through each step: variable definition, constraint identification, and the objective function. Although the final response must remain in the compact form, this strategy tends to generate more thorough formulations. The prompts used for candidate solution generation are given in Appendix B.

\subsection{Ranking and Selecting the Final Formulation}
\label{subsec:ranking}
Although each $f_i(T)$ may include a valid set of decision variables, constraints, and an objective function, certain candidates can more accurately reflect the requirements.
To address this variability, our methodology proposes a ranking framework that systematically compares multiple candidates rather than evaluating them in isolation.
Each candidate formulation is evaluated for its completeness, correctness, and logical consistency through a structured comparative evaluation process using an instruction-tuned LLM.
Additionally, the formatting of the mathematical model is evaluated to ensure adherence to standard conventions for LP, ILP, or MILP formulations. Unlike traditional ranking mechanisms that rank all candidate solutions in a single step, our methodology adopts a sample-by-sample ranking strategy, where candidates are compared in pairs.
For example, if there are three candidates $f_1(T)$, $f_2(T)$, and $f_3(T)$, we first compare $f_1(T)$ and $f_2(T)$, selecting the better of the two, and then compare the best candidate with $f_3(T)$. While this approach requires one more LLM inference than single-step ranking, our empirical study shows that it provides higher and more reliable solution quality.

One of the critical aspects of our proposed pipeline is the requirement for fully automated inference and evaluation, implying that the determination of the best candidate solution should proceed without external annotations or human involvement.
When we treat evaluation as a downstream task, we must provide or produce data in a structured format to make the process more effective.
Our initial tests with open-source LLMs indicate that the ranking model's responses do not consistently follow a single pattern across different samples---even when the sampling temperature is set to a very low value.
This presents an intriguing paradox: LLMs inherently generate unstructured text, reflecting their training on massive, mostly unstructured datasets.
To achieve formatted responses, we propose a structured output-parsing approach, where each LLM inference is accompanied by a predefined data model (class definition) that helps extract the final best candidate formulation. In our data model, we included a field to denote the best solution (either 1 or 2) for pairwise comparisons, alongside a second field for the second-best choice. For example, if the model identifies candidate~$2$ as the best, it must label it accordingly (best = $2$, second-best = $1$) to prevent random or incorrect outputs. While this rigid formatting may seem trivial to a human annotator---given that if one solution is better, the other must logically be second-best---it proved highly effective in reducing hallucinations.
However, a separate study by Tam~et~al. \cite{tam-etal-2024-speak} presents empirical evidence that LLMs face difficulties with reasoning tasks when subject to format constraints. Moreover, their findings show that the models’ reasoning performance deteriorates further as these constraints become stricter. In our preliminary experiments, we observed a similar trend, especially among smaller LLMs: while the restrictions effectively prevented any ``hallucinations” (the model consistently provided only “1” or “2” and reversed them for best and second-best), they also caused the model to frequently pick the simplest candidate solution. This usually meant selecting the direct prompt query rather than the few-shot or chain-of-thought versions as the best option. We hypothesize that these rigorous formatting requirements may hinder the model’s deeper reasoning processes, prompting it to default to the simplest outcome.
According to OpenAI \cite{openai_structured_outputs_api_2024}, the quality of an LLM's final output can be enhanced by incorporating reasoning steps within the schema. This field allows the model to explicitly outline its reasoning process before providing the final answer in a separate field. Building on this insight, we added a `reasoning\_steps' text field to our data model, instructing the LLM to justify choosing one response over another as shown in the following data model. This modification led to an improvement in performance.
\begin{lstlisting}[style=mypython]
class ComparisonResult(BaseModel):
    best: int  # 1 for Candidate 1, 2 for Candidate 2
    second_best: int  # 1 for Candidate 1, 2 for Candidate 2
    reasoning_steps: str  # Reason for ranking
\end{lstlisting}

\noindent The prompt used for ranking the solutions is given in Appendix C.

\section{Evaluation Metrics}
\label{sec:Evaluation}
We evaluate the similarity between the generated and reference formulations using a combination of metrics. We use two overlap-based metrics: BLEU \cite{papineni-etal-2002-bleu}, a precision-oriented metric that measures token n-gram overlap and is widely used in machine translation, and ROUGE-1, ROUGE-2, and ROUGE-L \cite{lin-2004-rouge}, recall-oriented metrics commonly applied in summarization tasks. Additionally, we utilize BERTScore \cite{Zhang*2020BERTScore:}, a representation-based metric that calculates cosine similarity between contextualized token embeddings and has demonstrated a stronger correlation with human judgments compared to BLEU and ROUGE across various tasks.

\subsection{Math-aware BERTScore}
The original BERT \cite{devlin-etal-2019-bert}, designed for general natural language tasks, struggles with mathematical equations due to its pre-training on corpora deficient in mathematical notation and a tokenizer not suited for LaTeX equation format. It cannot effectively process symbols or capture the structural nuances of mathematical expressions. Constructs like `\textbackslash sum', `\textbackslash frac', or `\textbackslash alpha' are fragmented into multiple tokens, disrupting their semantic meaning as unified entities. Since mathematical equations rely heavily on precise relationships and syntactic structure, this fragmentation compromises coherence and leads to significant information loss. We use embeddings from Math-aware BERT \cite{Reusch2022TransformerEncoderAD} to address this limitation for cosine similarity in BERTScore. This model, pre-trained on the MathSE dataset and extended tokenizer with $501$ LaTeX tokens, effectively handles mathematical expressions. Fine-tuning pairs questions with answers using ARQMath annotations and formulas are processed as Symbol Layout and Operator Trees. These adaptations improve the baseline BERT’s performance in mathematical content retrieval.

However, existing metrics for comparing mathematical formulations have key limitations. BLEU and ROUGE rely on exact token matching, missing semantic equivalence in structurally different but equivalent formulations. Math-aware BERTScore addresses this but lacks a mechanism for capturing nuanced equivalence in complex constructs, emphasizing the need for a more semantically aligned metric for mathematical content. In a similar task, Manas et al. \cite{Mañas_Krojer_Agrawal_2024} proposed an instruction-tuned LLM-based LAVE metric to evaluate open-ended visual question answering tasks by including semantic reasoning and contextual understanding. By aligning with human judgment and addressing gaps in traditional metrics, LAVE demonstrated significant improvements in evaluating nuanced and complex answers. Motivated by these advancements, we propose LLM-Assisted Mathematical Evaluation Score (LAME), an automated evaluation framework that evaluates the quality of mathematical formulations generated for LP, ILP, or MILP problems.

\subsection{LAME Score}
 Each evaluation example comprises a problem description \( p \), a ground truth mathematical formulation \( g \), and a candidate mathematical formulation \( c \). The objective of LAME is to evaluate the candidate formulation \( c \) by comparing it with the ground truth \( g \) and considering the context provided by the problem description \( p \). To achieve this, we utilize the in-context learning capabilities of instruction-tuned LLMs. We construct a textual prompt using \( p \), \( g \), and \( c \), which is then fed to the LLM to generate a detailed evaluation across multiple criteria, including the correctness of the objective function, accuracy of constraints, definition of variables, and overall validity of the formulation. Below, we describe the design decisions underlying LAME scoring.

\subsubsection{Choosing a Large Language Model}
The selection of a suitable LLM is critical to the performance of LAME, as its effectiveness depends upon the model's ability to analyze and compare mathematical formulations. We pose the evaluation as a close-ended assessment task with detailed scoring, which aligns well with the reasoning capabilities of instruction-tuned LLMs. Consequently, we select the Flan-T5 model family \cite{JMLR:v25:23-0870} as the primary LLM for LAME. This choice is informed by the model's instruction-tuning and chain-of-thought reasoning capabilities, which enable it to provide step-by-step evaluations for complex tasks.

To validate the generalizability of LAME, we design it to work with multiple LLMs. In addition to Flan-T5, we test LAME using other instruction-tuned models. While Flan-T5 serves as the baseline, this adaptability ensures that LAME can incorporate future advancements in LLMs with minimal changes to its structure.

\subsubsection{Prompt for Mathematical Formulation Evaluation}
The prompt for the LAME score comprises three main components: a task definition, demonstrations of evaluations, and the input test example. The task definition specifies the evaluation criteria, including:
\begin{enumerate}
\renewcommand{\labelenumi}{\roman{enumi}.}
    \item Objective Function Correctness: Is the objective function correctly formulated to match the problem description? (Score out of 10)
    \item Constraint Accuracy: Are the constraints comprehensive and correctly stated? (Score out of 10)
    \item Variable Definitions: Are the decision variables properly defined and utilized? (Score out of 10)
    \item Overall Validity: Does the formulation faithfully represent the problem requirements? (Score out of 40)
\end{enumerate}

Each demonstration includes a problem description, ground truth formulation, candidate formulation, and the corresponding evaluation output with detailed scores and rationales. The demonstrations are curated to cover diverse cases, including examples with varying complexity, precision, and types of errors. Users can extend these demonstrations to adapt the framework to specific domains. The test example consists of a problem description, ground truth formulation, and candidate formulation. The LLM generates an evaluation in the format:

\begin{lstlisting}
Objective Function Correctness: X/10
Constraint Accuracy: Y/10
Variable Definitions: Z/10
Overall Score: W/40
Overall Analysis: [A brief summary of the evaluation]
\end{lstlisting}
The prompt also includes filtering mechanisms for references and rationalization. By default, outlier references are not filtered, as all components in a mathematical formulation are typically crucial. Rationalization—providing a justification for each score—is an integral part of the task to improve explainability and consistency.

\subsubsection{Scoring Function}
The scoring function extracts numerical ratings from the LLM-generated text, processes them, and aggregates them into a final evaluation score. Using regular expressions, it identifies specific patterns in the generated text to extract ratings for each criterion and the overall score. In cases where a score is missing or ambiguous, default values are applied to maintain consistency. This ensures that LAME can handle noisy or unexpected outputs without compromising reliability. Each score is normalized to a range of 0 to 1, and the overall score is computed as a weighted sum of these normalized values:
\[
S = w_o \cdot S_o + w_c \cdot S_c + w_v \cdot S_v + w_a \cdot S_a
\]
where \( S_o, S_c, S_v, \) and \( S_a \) represent the normalized scores for the objective function, constraints, variables, and overall validity, respectively. The weights \( w_o, w_c, w_v, w_a \) are predefined to reflect the importance of each criterion (e.g., \( w_o = 0.4, w_c = 0.3, w_v = 0.2, w_a = 0.1 \)).

\section{Findings and Analysis}
\label{sec:Findings}
\subsection{Technical Details}
The evaluation was conducted on a server equipped with an NVIDIA H100 NVL GPU featuring 96GB VRAM, 64GB RAM, and an Intel Xeon Gold CPU with 64 cores and 128 threads. As the focus was on evaluating the in-context learning capabilities of the LLMs, no fine-tuning was applied during the evaluation process.
\subsection{Results}
We present the baseline LLM performance on the curated NL4RA dataset in Table \ref{tab:model_performance}, where we report overlap based metrics such as BLEU, ROUGE 1, ROUGE 2, ROUGE L, BERTScore, and our proposed LAME 1, LAME 3, and LAME 5 metrics. The evaluation includes four language models, namely Llama 3.1 at two parameter scales (70B and 8B) and Phi 3 at two parameter scales (3.5B and 14B), with the temperature parameter varied among 0, 0.5, and 1. 

An interesting observation from the ROUGE and BLEU scores is that they remain consistently low across all temperature settings and models. This result supports our initial hypothesis that exact token matching is not suitable for comparing mathematical expressions, as notational differences or symbol placement can cause low n-gram overlaps. For instance, despite Llama-3.1-8B at temperature $0$ achieving moderately higher ROUGE-1 ($0.4534$) relative to Phi-3-14B ($0.4321$), its BLEU score is still below 0.06. Such discrepancies reveal the inherent weakness of purely lexical measures, as small notational variations in LaTeX code can escape direct matching. BERTScore, adapted to math-aware embeddings, tries to address these issues by focusing on contextual meaning rather than exact token matches. Nevertheless, as Table \ref{tab:model_performance} shows, the BERTScore values across most configurations are relatively close (ranging mostly between $0.70$ and $0.79$) and do not fully correlate with human evaluations. Our analysis suggests that large formulations with more elaborate LaTeX equations can confuse the tokenizer of standard BERT or even math-aware BERT, causing suboptimal alignment and limited variation in BERTScore. The frequent mis-tokenization of domain-specific variables and constraint names further lessens the reliability of this metric in ranking the best formulation.  

To address these weaknesses, we rely on our LAME scores, shown in the last three columns of Table \ref{tab:model_performance}. LAME-1, LAME-3, and LAME-5 each use an instruction-tuned LLM to judge how well a candidate formulation aligns with a ground truth while conditioning on $1$, $3$, or $5$ examples, respectively, in an in-context learning scenario. Across all models, LAME-\{3,5\} exhibits a wider score spread than BERTScore or ROUGE, indicating that the method captures a richer spectrum of correctness criteria. For instance, Phi-3.5-3.8B at temperature $1.0$ achieves a BERTScore of $0.7092$, but obtains a notably low LAME-5 of $0.226$. Moreover, comparing LAME-1, LAME-3, and LAME-5 reveals that LAME-5 consistently demonstrates the highest correlation with manual inspections of the generated formulations. According to our partial human annotations of the LLM-generated responses, LAME-5 effectively penalizes these incorrect expansions while rewarding correct variable usage and coherent constraint sets. By contrast, LAME-1 sometimes oversimplifies the evaluation, leading to less sensitive gradations among candidate solutions. These findings are consistent with our prior analysis that purely lexical or token-level metrics can be misaligned with human judgment, especially for syntactically complex LaTeX formulations. In several instances, BERTScore and ROUGE underreported differences between models that humans identified as significant; likewise, BLEU remained too low to offer meaningful distinctions. By contrast, LAME-\{3,5\} showed an enhanced capacity.

\begin{table*}[h]
\scriptsize
\centering
\caption{Performance Metrics for Various Models Across Different Temperatures}
\label{tab:model_performance}
\begin{tabular}{|c|c|c|c|c|c|c|c|c|c|}
\hline
\textbf{Temp.} & \textbf{Model}          & \textbf{BERT Score} & \textbf{ROUGE-1} & \textbf{ROUGE-2} & \textbf{ROUGE-L} & \textbf{BLEU} & \textbf{LAME-3} & \textbf{LAME-5} & \textbf{LAME-1} \\ \hline
0                    & Llama-3.1-70B           & 0.7559              & 0.4249           & 0.1624           & 0.2797           & 0.0326        & 0.7728          & 0.7492          & 0.89            \\ \hline
0                    & Llama-3.1-8B            & 0.7873              & 0.4534           & 0.1684           & 0.2758           & 0.0508        & 0.7306          & 0.7152          & 0.854           \\ \hline
0                    & Phi-3.5-3.8B            & 0.7181              & 0.1166           & 0.0077           & 0.0723           & 0.0014        & 0.2016          & 0.202           & 0.336           \\ \hline
0                    & Phi-3-14B               & 0.7458              & 0.4321           & 0.1341           & 0.2006           & 0.0263        & 0.6802          & 0.6504          & 0.83            \\ \hline
0.5                  & Llama-3.1-70B           & 0.7727              & 0.4465           & 0.1683           & 0.2806           & 0.0325        & 0.7782          & 0.7451          & 0.91            \\ \hline
0.5                  & Llama-3.1-8B            & 0.7817              & 0.4534           & 0.1689           & 0.2645           & 0.0439        & 0.7444          & 0.7088          & 0.87            \\ \hline
0.5                  & Phi-3.5-3.8B            & 0.7758              & 0.3684           & 0.0938           & 0.1851           & 0.0112        & 0.6330          & 0.5738          & 0.764           \\ \hline
0.5                  & Phi-3-14B               & 0.7409              & 0.4286           & 0.1328           & 0.2068           & 0.0282        & 0.7018          & 0.6698          & 0.864           \\ \hline
1                    & Llama-3.1-70B           & 0.7830              & 0.4345           & 0.1463           & 0.2656           & 0.0291        & 0.7621          & 0.713           & 0.914           \\ \hline
1                    & Llama-3.1-8B            & 0.7584              & 0.4389           & 0.1574           & 0.2636           & 0.0413        & 0.7572          & 0.7152          & 0.872           \\ \hline
1                    & Phi-3.5-3.8B            & 0.7092              & 0.0979           & 0.0023           & 0.0587           & 0.0010        & 0.1498          & 0.226           & 0.254           \\ \hline
1                    & Phi-3-14B               & 0.7821              & 0.4126           & 0.1119           & 0.1935           & 0.0196        & 0.7360          & 0.6862          & 0.864           \\ \hline
\end{tabular}

\end{table*}

\begin{figure}
    \centering
    \includegraphics[width=0.7\columnwidth]{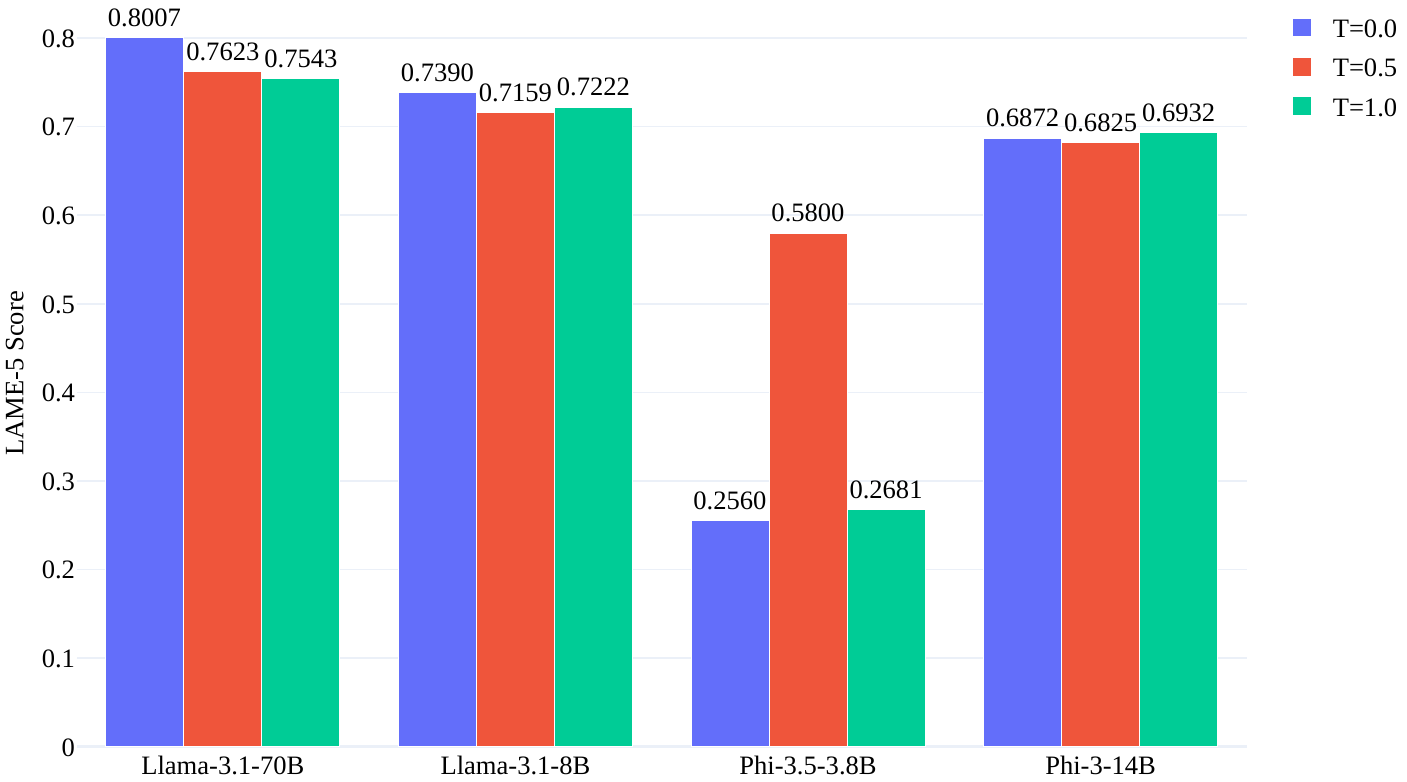}
    \caption{LAME-5 scores for different models at various temperatures using the proposed framework}
    \label{fig:LAME-5_scores}
\end{figure}

\figureautorefname~\ref{fig:LAME-5_scores} presents the LAME-5 performance of our proposed multi-candidate LLM framework, LM4Opt-RA, applied to four models at three temperature settings. Compared to our baseline results in Table \ref{tab:model_performance}, there is an improvement in LAME-5 scores under the new approach, particularly for Llama-3.1-70B at \(T=0.0\), which now exceeds $0.80$. Although Llama-3.1-70B still shows a slight decrease in performance when the temperature is raised to 1.0, its LAME-5 scores remain robust relative to the baseline. Interestingly, Phi-3.5-3.8B exhibits sharper fluctuations at \(T=0.0\), \(T=0.5\) and \(T=1.0\), which shows the model’s sensitivity to randomness in generation.

\subsection{Effect of Prompting Strategies on Solution Complexity}
\label{subsec:prompting_complexity}

\begin{figure}
    \centering
    \includegraphics[width=0.7\columnwidth]{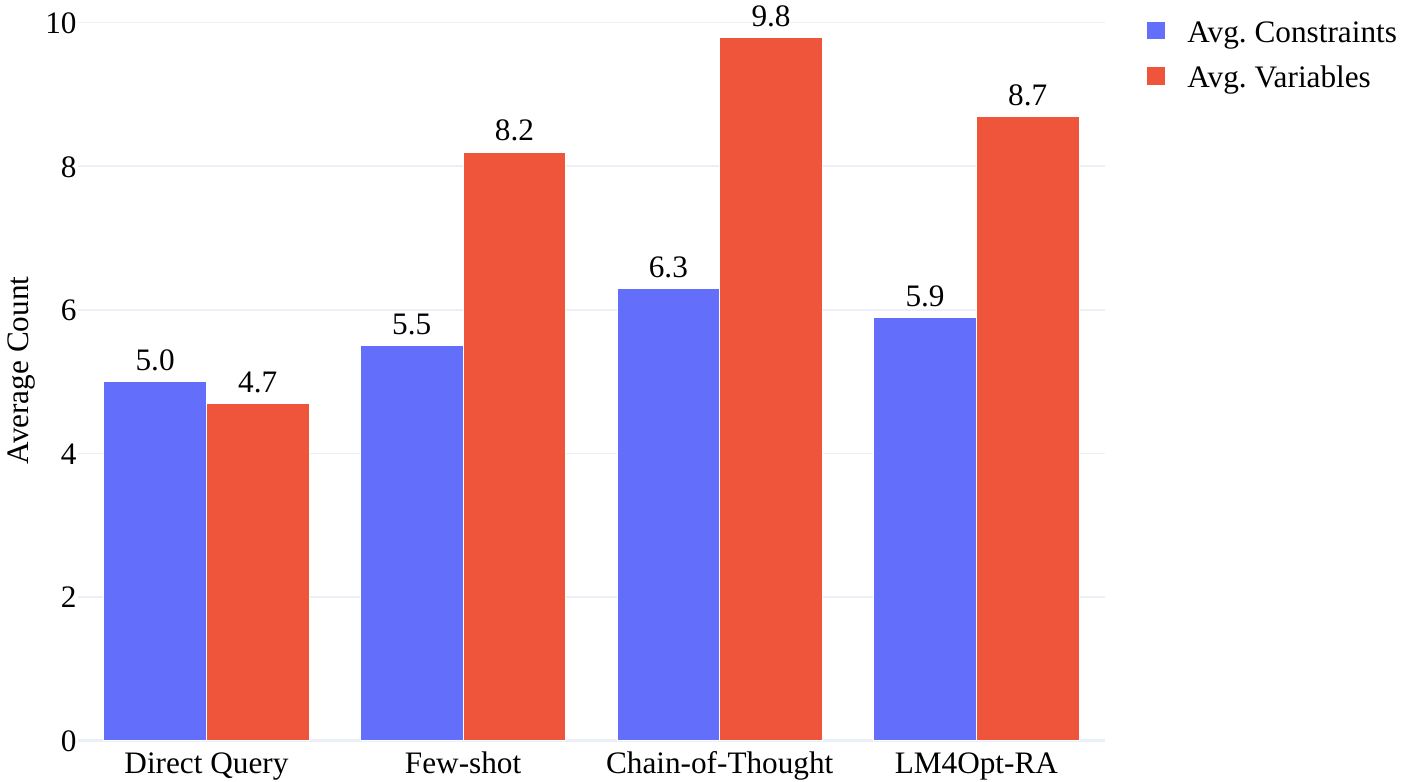}
    \caption{Comparison of average number of constraints and variables across different strategies}
    \label{fig:avg_con_var}
\end{figure}

\begin{figure*}
    \centering
    \includegraphics[width=\textwidth]{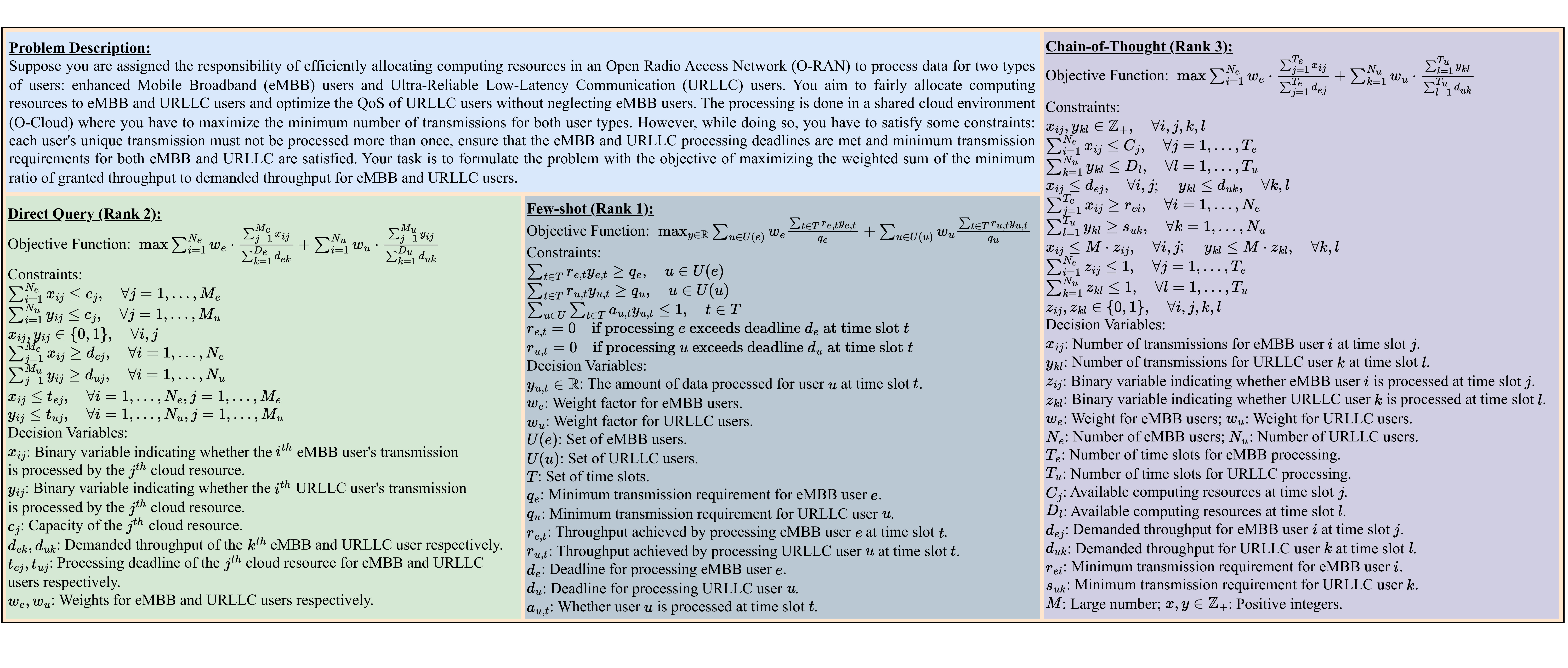}
    \caption{Side-by-side comparison of three candidate solutions (Direct Query, Few-Shot, and Chain-of-Thought) for the same resource allocation problem description}
    \label{fig:exmp}
\end{figure*}

Figure~\ref{fig:avg_con_var} compares the average number of constraints and variables generated by each prompting strategy: direct query, chain-of-thought, few-shot, and our proposed LM4Opt-RA. A notable observation is that the direct query approach generates the simplest formulations: the fewest constraints and a smaller number of variables. Upon manual review, these solutions often fail to capture the full intricacy of the problem description, overlooking important constraints or variables. In contrast, the chain-of-thought strategy produces the largest average counts. While this approach generally provides better coverage and accounts for more aspects of the problem, it can also become excessively complex—introducing additional variables or constraints not strictly required by the original scenario. The few-shot strategy stands somewhere in between, occasionally including more variables but fewer constraints. Our proposed multi-candidate framework, LM4Opt-RA, selects the best solution among these three strategies. Consequently, the average constraints and variables for LM4Opt-RA are closer to the original formulations in the curated dataset. Moreover, manual inspections reveal that LM4Opt-RA most frequently picks either a few-shot or chain-of-thought solution, indicating that these strategies often provide a more comprehensive alignment with the real-world complexity of network resource allocation problems.
To illustrate how each strategy handles the same problem description, Figure~\ref{fig:exmp} presents three candidate formulations side by side---along with their final rankings. We observe that the direct query solution omits important details and yields a more compact formulation, whereas the chain-of-thought version introduces additional complexity and sometimes unnecessary variables. By contrast, the few-shot solution balances more effectively between completeness and clarity, resulting in its top ranking by our framework.

\subsection{Effect of Formatting Constraints on Reasoning Capabilities}
We detailed our candidate solution ranking mechanism in \textbf{Ranking and Selecting the Final Formulation}, arriving at the final proposed methodology through a step-by-step process, moving from lenient to stricter formatting constraints. Initially, we used generic prompts where the LLM was asked to rank solutions as best, second-best, and third-best. While the LLM performed satisfactorily, identifying discrepancies with the ground truth reasonably well, its responses were overly descriptive and lacked a consistent format across samples. 
To address this, we implemented stricter formatting, instructing the LLM to provide only the ranks without additional text. Most models in the Llama series handled this task effectively, although the smaller Llama model often ranked the easiest and smallest solution as the best, even when it did not align well with the problem description which indicates a notable decrease in reasoning capability. The Phi models, particularly the smaller one, failed noticeably to adhere to the specified format. Instead, the smaller Phi model frequently produced irrelevant, overly verbose gibberish responses. This issue arose due to the lengthy prompt, which included the problem description, three candidate solutions, and detailed ranking and formatting instructions. Although the prompt length fell within the model's context limit, the smaller Phi model exhibited unpredictable behavior with longer inputs. Consequently, the shift from generic conversational prompts to strict formatting diminished the LLM's reasoning capabilities.
To ensure structured responses, we then adopted a structured LLM output with a specific data model, and all models provided rankings without extra text. However, we observed that most rankings followed the sequence \{1, 2, 3\}, leading us to hypothesize that the strict response format reduced the model's reasoning abilities, prompting it to default to selecting the simplest solution as the best. At this stage, we incorporated reasoning steps into the data model, which significantly improved the outcomes and enhanced the LLM's reasoning capabilities.

\subsection{Limitations}
\subsubsection{Academic Sources for Problem Descriptions}
We acknowledge that the language used in the NL4RA dataset is predominantly formal and domain-specific, owing to the peer-reviewed scientific articles from which these problem statements are derived. While this choice allows us to focus on the technical precision and consistency necessary for network resource allocation tasks, it may not fully reflect more colloquial or layperson-written descriptions encountered in everyday contexts. Consequently, the framework’s performance could vary when adapting to problem statements that lack specialized terminology or use more casual, non-technical language---an area that requires further exploration.
\subsubsection{Reliance on Automated Evaluation Metrics}
Although the LAME score provides a math-aware assessment of LP, ILP, and MILP formulations, it cannot fully replicate the nuanced judgment of domain experts, particularly regarding problem-specific constraints or specialized terminology. Subtle inaccuracies may consequently slip through which emphasizes the need for human oversight in high-stakes or sensitive environments. However, by substantially reducing the resource requirements for preliminary evaluations, LAME broadens access to advanced optimization techniques, making them more approachable for a wider range of users and facilitating quicker, more iterative problem-solving processes.
\subsubsection{Choice of Open-Source Models}
Our empirical evaluation focuses on the open-source Llama and Phi language models, ensuring that the experiments are reproducible and transparent. We have chosen these LLMs as various benchmarks recognize them among the most capable open-source LLMs for reasoning tasks. However, this choice may not fully represent the capabilities of more advanced or proprietary systems, such as GPT-\{4, 4o, o1\}. Future investigations examining a broader array of models could yield a more comprehensive understanding of performance and generalizability in the domain of resource allocation.

\section{Conclusion}
\label{sec:Conclusions}
We introduce NL4RA, the first curated dataset explicitly tailored for network resource allocation, creating a foundational benchmark that captures the dynamic, heterogeneous constraints of real-world problems and supports future research in this domain.
In addition, we have presented LM4Opt-RA, a multi-candidate LLM framework for automatically generating mathematical formulations of network resource allocation problems. By implementing direct, few-shot, and chain-of-thought prompts, followed by a ranking mechanism, we show that LM4Opt-RA produces more coherent, complete, and context-aligned formulations than single-prompt approaches. Our evaluation with standard overlap-based metrics (BLEU, ROUGE) highlights their inadequacy for capturing nuanced, notation-heavy mathematical expressions, while math-aware BERTScore only partially addresses these limitations. In contrast, our proposed LAME metric, particularly LAME 5, shows a stronger correspondence with human evaluations, highlighting its promise as a reliable automated judge of formulation correctness.
Looking ahead, future work can expand LM4Opt-RA to broader classes of resource allocation problems, including multi-objective or stochastic formulations, and investigate advanced embedding strategies that better handle domain-specific symbolic expressions. We expect these refinements to further narrow the gap between machine-generated formulations and the expertise of human domain specialists.

\bibliographystyle{unsrt}  
\bibliography{sample}

\appendix

\section{Funding}
This research was supported by the \textbf{Natural Sciences and Engineering Research Council of Canada (NSERC)}.

\section{Selected Works for NL4RA}
\label{sec:selected_works_NL4RA}

  NL4RA consists of various resource allocation optimization problems in the networking domain. Table \ref{Tab-SelectedPapers-NL4RA} categorizes the selected works based on multiple network domains relevant to the NL4RA dataset. Furthermore, Table \ref{Tab-SampleMapping-NL4RA} presents a mapping between selected works and specific problem instances from the NL4RA dataset.

\renewcommand{\thetable}{A.\arabic{table}}
\setcounter{table}{0}
\begin{table}[htbp]
\centering
\caption{Categorization of Selected Works based on Different Network Domains for NL4RA}
\label{Tab-SelectedPapers-NL4RA}
\begin{tabular}{|p{7cm}|p{9.5cm}|}

\hline
\textbf{Network Domain} & \textbf{References} \\
\hline
Radio Access Networks (Cloud-RAN, MIMO, HetNet, Cognitive Radio Network and 5G NR) & \cite{salameh2022opportunistic}, \cite{hadi2020patient},\cite{sharara2022policy},\cite{javad2021re},\cite{zhai2017adaptive},\cite{feng2017adaptive}, \cite{saghezchi2017energy}, \cite{xie2017energy}\\
\hline
Network Technologies (Network Slicing, NFV and SDN) & \cite{you2018resource}, \cite{debbabi2022inter},\cite{emu2021ensemble}, \cite{fayad20235g},\cite{pan2020multi},\cite{javad2021re}, \cite{yang2020towards}, \cite{ko2022pdras}, \cite{yang2019two},\cite{fendt2018network},\cite{zhai2017adaptive},\cite{8204056}, \cite{xie2017energy}, \cite{jia2017efficient}, \cite{spinelli2022edge}, \cite{zoha2017leveraging},\cite{gao2019deep},\cite{sattar2019optimal}, \cite{yu2019network},\cite{de2020optimal}, \cite{9929619}, \cite{9827120}, \cite{almasaeid2023minimum}\\
\hline
Advanced Distributed Systems (Distributed Cloud, F-RAN, Edge Computing, Fog Computing, and UAV Relay Network) & \cite{alam2021resource},\cite{emu2021ensemble},\cite{alsheyab2019near},\cite{de2020optimal},\cite{spinelli2022migration}, \cite{kim2022modems}, \cite{pan2020multi},\cite{javad2021re}, \cite{spinelli2022edge},\cite{gao2019deep}, \cite{mharsi2018joint}, \cite{de2020optimal}\\
\hline
Communication Technologies (D2D and Wireless Communication) & \cite{al2021joint},\cite{hassan2017interference}, \cite{hussain2018system},\cite{hossen2018relax}, \cite{hassan2019online}, \cite{shokrnezhad2018joint}, \cite{you2017load},\cite{vlachos2017moca}, \cite{9068264}, \cite{alnakhli2024optimizing}\\
\hline
\end{tabular}
\end{table}

\begin{table*}[htbp]
\scriptsize
\centering
\caption{Selected Paper Vs Sample Dataset Mapping - NL4RA}
\label{Tab-SampleMapping-NL4RA}
\begin{tabular}{|>{\raggedright\arraybackslash}m{13cm}|>{\raggedright\arraybackslash}m{3.5cm}|}
\hline
\textbf{Selected Works} & \textbf{Problem Instance} \\
\hline
Resource Optimization With Flexible Numerology and Frame Structure for Heterogeneous Services \cite{you2018resource} & Sample 1\\
\hline
A Joint Algorithm for Resource Allocation in D2D 5G Wireless Networks \cite{al2021joint} & Sample 2, Sample 3\\
\hline
Inter-slice B5G Bandwidth Resource Allocation \cite{debbabi2022inter} & Sample 4\\
\hline
A Resource Allocation Model Based on Trust
Evaluation in Multi-Cloud Environments \cite{alam2021resource}  & Sample 5\\
\hline
Ensemble Deep Learning Assisted VNF
Deployment Strategy for Next-Generation
IoT Services \cite{emu2021ensemble}  & Sample 6\\
\hline
Interference Minimization in D2D Communication Underlaying Cellular Networks \cite{hassan2017interference} & Sample 7\\
\hline
Near-Optimal Resource Allocation Algorithms for
5G+ Cellular Networks \cite{alsheyab2019near} & Sample 8\\
\hline
System Capacity Maximization With Efficient
Resource Allocation Algorithms in
D2D Communication \cite{hussain2018system} & Sample 9\\
\hline
Relax online resource allocation algorithms for D2D communication \cite{hossen2018relax} & Sample 10\\
\hline
 An online resource allocation algorithm to minimize system
interference for inband underlay D2D communications \cite{hassan2019online} & Sample 11\\
\hline
Optimal Allocation of vBBUs Considering Distance
Between MDC and RRH in F-RANs \cite{de2020optimal} & Sample 12\\
\hline
5G Millimeter Wave Network Optimization: Dual
Connectivity and Power Allocation Strategy \cite{fayad20235g} & Sample 13, Sample 14, Sample 15\\
\hline
A Migration Path Toward Green Edge Gaming \cite{spinelli2022migration} & Sample 16\\
\hline
MoDEMS: Optimizing Edge Computing Migrations
for User Mobility \cite{kim2022modems} & Sample 17\\
\hline
Opportunistic non-contiguous OFDMA scheduling framework for
future B5G/6G cellular networks \cite{salameh2022opportunistic} & Sample 18\\
\hline
Patient-Centric HetNets Powered by Machine
Learning and Big Data Analytics for 6G Networks \cite{hadi2020patient} & Sample 19\\
\hline
A Multi-Dimensional Resource Crowdsourcing
Framework for Mobile Edge Computing \cite{pan2020multi} & Sample 20\\
\hline
Policy-Gradient-Based Reinforcement Learning for
Computing Resources Allocation in O-RAN \cite{sharara2022policy} & Sample 21\\
\hline
Re-configuration of UAV Relays in 6G Networks \cite{javad2021re} & Sample 22\\
\hline
Towards 6G Joint HAPS-MEC-Cloud 3C Resource Allocation for Delay-Aware Computation Offloading \cite{yang2020towards} & Sample 23\\
\hline
PDRAS: Priority‑Based Dynamic Resource Allocation
Scheme in 5G Network Slicing \cite{ko2022pdras} & Sample 24\\
\hline
Two-Tier Resource Allocation in Dynamic Network
Slicing Paradigm with Deep Reinforcement
Learning \cite{yang2019two} & Sample 25\\
\hline
A Network Slice Resource Allocation and
Optimization Model for End-to-End Mobile Networks \cite{fendt2018network} & Sample 26\\
\hline
Adaptive Codebook Design and Assignment for
Energy Saving in SCMA Networks \cite{zhai2017adaptive} & Sample 27, Sample 28\\
\hline
Adaptive Pilot Design for Massive MIMO HetNets
with Wireless Backhaul \cite{feng2017adaptive} & Sample 29\\
\hline
Economic Node Allocation in Software Defined
Wireless Networks with Forecasted Traffic and
Distance Constraints \cite{8204056} & Sample 30\\
\hline
Energy-aware relay selection in cooperative wireless networks: An
assignment game approach \cite{saghezchi2017energy} & Sample 31\\
\hline
Energy-Efficient Joint Content Caching and Small
Base Station Activation Mechanism Design in Heterogeneous Cellular Networks \cite{xie2017energy} & Sample 32\\
\hline
Efficient caching resource allocation for
network slicing in 5G core network \cite{jia2017efficient} & Sample 33\\
\hline
Joint power control and channel assignment in uplink IoT Networks: A noncooperative game and auction based approach \cite{shokrnezhad2018joint} & Sample 34\\
\hline
Edge Gaming: A Greening Perspective \cite{spinelli2022edge} & Sample 35\\
\hline
Leveraging Intelligence from Network CDR
Data for Interference Aware Energy
Consumption Minimization \cite{zoha2017leveraging} & Sample 36\\
\hline
Load Optimization With User Association
in Cooperative and Load-Coupled
LTE Networks \cite{you2017load} & Sample 37, Sample 38\\
\hline
MOCA: Multiobjective Cell Association for Device-to-Device Communications \cite{vlachos2017moca} & Sample 39, Sample 40\\
\hline
Deep Neural Network Task Partitioning and
Offloading for Mobile Edge Computing \cite{gao2019deep} & Sample 41\\
\hline
Joint Optimization of Communication Latency and
Resource Allocation in Cloud Radio Access Networks \cite{mharsi2018joint} & Sample 42\\
\hline
Distributed Radio Slice Allocation in Wireless
Network Virtualization: Matching
Theory Meets Auctions \cite{9068264} & Sample 43\\
\hline
Optimal Slice Allocation in 5G Core Networks \cite{sattar2019optimal} & Sample 44\\
\hline
Network Function Virtualization Resource Allocation
Based on Joint Benders Decomposition and ADMM \cite{yu2019network} & Sample 45\\
\hline
Optimal Virtual Network Function Deployment
for 5G Network Slicing in a Hybrid
Cloud Infrastructure \cite{de2020optimal} & Sample 46\\
\hline
RAN Slice Access Control Scheme Based on
QoS and Maximum Network Utility \cite{9929619} & Sample 47\\
\hline
Slice-Aware Resource Calendaring in Cloud-based Radio Access Networks \cite{9827120} & Sample 48\\
\hline
Optimizing spectrum efficiency in 6G
multi-UAV networks through source correlation
exploitation \cite{alnakhli2024optimizing} & Sample 49\\
\hline
Minimum cost spectrum allocation with QoS guarantees in multi-interface multi-hop dynamic spectrum access networks \cite{almasaeid2023minimum} & Sample 50\\
\hline
\end{tabular}
\end{table*}

\section{Prompt Templates for Candidate Generation}
\label{sec:prompt_templates}

To systematically generate mathematical formulations for network resource allocation optimization problems, we designed three distinct prompt strategies: Direct Prompt, Few-Shot Prompt, and Chain-of-Thought Prompt. The detailed descriptions of these prompts are presented below.

\subsection*{Direct Prompt}
The Direct Prompt focuses on providing minimal instructions to the LLM for translating a natural language problem description into a mathematical formulation. The prompt emphasizes conciseness, with strict adherence to LaTeX formatting for all equations and definitions, as shown below:

\begin{lstlisting}[style=promptblock]
You are an expert in writing mathematical formulations for network resource allocation optimization problems. I will give you a problem description. Your task is to give me a problem formulation that includes the optimization model based on the decision variables, constraints, and objective functions from the problem description.

Your response MUST contain only the problem formulation and equations or mathematical terms MUST be in Latex format. After the objective function and constraints, include decision variable definitions as well. DO NOT add additional text, or explanation before or after it.

Problem description to formulate: 
\end{lstlisting}

\subsection*{Few-Shot Prompt}
The Few-Shot Prompt provides an example problem and its corresponding mathematical formulation to guide the LLM. The example problem and its formulation are embedded within the prompt, as demonstrated below:

\begin{lstlisting}[style=promptblock]
You are an expert in writing mathematical formulations for network resource allocation optimization problems. I will give you a problem description. Your task is to give me a problem formulation that includes the optimization model based on the decision variables, constraints, and objective functions from the problem description.
Your response MUST contain only the problem formulation and equations or mathematical terms MUST be in Latex format. After the objective function and constraints, include decision variable definitions as well. DO NOT add additional text, or explanation before or after it. Here is an example:
----------------------------------
"Suppose a base station has two categories of services, denoted by $K(\ell)$ and $K(c)$, where services of $K(\ell)$ are prioritized over $K(c)$. For any service $k \in K(\ell)$, the data demand is denoted by $q_k$ (in bits) and must be met with a latency tolerance (time until data has been fully transmitted) of $\tau_k$. 
The target is to maximize the total throughput of $K(c)$ through optimal resource configuration of numerology and frame structure for each service, subject to latency and demand constraints for $K(\ell)$. You can consider the resource configuration of numerology and frame structure as blocks, and define a candidate set $B$ for blocks and a set of basic units $I$ where each unit i is associated with each block to ensure that the services are non-overlapping. For each $b \in B$, the achieved throughput on block $b$, if $b$ is assigned to service $k$ ($k \in K$), is denoted by $r_{{b,k}}$.

The optimization task is to select the blocks for each service so that the latency and demand
requirements for $K(\ell)$ are met, without overlapping the chosen ones."
Example Problem Formulation:
\max_{{x \in \{{0,1\}}}} \sum_{{k \in K(c)}} \sum_{{b \in B}} r_{{b,k}} x_{{b,k}}
\\ \text{{s.t.}}
& \sum_{{b \in B}} r_{{b,k}} x_{{b,k}} \geq q_k, \quad k \in K(\ell) \\
& r_{{b,k}} = 0 \quad \text{{if block }} b \text{{ exceeds the latency }} \tau_k \text{{ for service }} k \in K(\ell) \\
& \sum_{{k \in K}} \sum_{{b \in B}} a_{{b,i}} x_{{b,k}} \leq 1, \quad i \in I
where:
- $x_{{b,k}} \in \{{0, 1\}}$: Whether block $b$ is assigned to service $k$. If $x_{{b,k}} = 1$, block $b$ is assigned to service $k$; otherwise, $x_{{b,k}} = 0$.
- $a_{{b,i}}$: Whether block $b$ includes basic unit $i$. $a_{{b,i}}=1$ if it includes the basic unit, otherwise 0.
- $K(\ell)$: Category of service that has strict latency requirement.
- $K(c)$: Another category of service that has a specific latency tolerance.
- $q_k$: Data demand (in bits) for any service $k \in K(\ell)$.
- $\tau_k$: Latency tolerance for service $k \in K(\ell)$.
- $B$: Set of blocks.
- $I$: Set of basic units.
- $r_{{b,k}}$: Achieved throughput on block $b$ if assigned to service $k$.
----------------------------------
Now, apply the same procedure for the following problem description:
\end{lstlisting}

\subsection*{Chain-of-Thought Prompt}
The chain-of-thought Prompt guides the LLM through a step-by-step reasoning process. The prompt is given below:

\begin{lstlisting}[style=promptblock]
You are an expert in mathematical optimization modeling for network resource allocation problems. Your task is to formulate optimization problems step by step by following these systematic reasoning stages:
1. Variable Identification:
   - Begin by carefully identifying all decision variables mentioned in the problem description.
   - For each variable, determine its domain (binary, continuous, integer) and define its indices and associated sets.
2. Constraint Analysis:
   - Systematically analyze the problem to identify resource limitations.
   - Note down technical requirements specific to the system.
   - Recognize and categorize system constraints.
3. Objective Function:
   - Identify the primary goal of optimization (maximize or minimize a specific metric).
   - Clearly express this goal as a mathematical formula.
   - Ensure the formulation aligns with the problem's stated objectives.
4. Mathematical Formulation:
   - Present a detailed mathematical representation:
     - Write the complete objective function using LaTeX.
     - List all constraints in LaTeX format.
     - Provide precise definitions for all variables and parameters.
Output Requirements:
- Your response MUST adhere to the following format, strictly using LaTeX for all mathematical
expressions:

\text{{Objective Function:}} & \quad <Objective in LaTeX> \\ 
\text{{s.t.}} \\
\text{{Constraints:}} & \quad <Constraints in LaTeX> \\
  where:
  - <Variable Definitions>
  - <Parameter Definitions>
  
Additional Requirements:
- Do NOT include any explanatory text or additional information outside this format.
- Your response should contain: 1. The objective function in LaTeX format.
2. All constraints in LaTeX format.
3. Complete definitions of variables and parameters.
Follow the format strictly and avoid any deviations. Problem description:
\end{lstlisting}

\section{Structured Ranking Prompt}
\label{sec:ranking_prompt}

The structured ranking prompt used for evaluating candidate solutions is given below:

\begin{lstlisting}[style=promptblock]
You are an expert in formulating and evaluating mathematical optimization problems. Below are two candidate solutions to a problem description.
1. **Read the problem description thoroughly** and carefully examine each candidate solution.
2. Evaluate them against these criteria:
   - **Completeness**: Does it include all necessary decision variables, constraints, and a proper objective function?
   - **Correctness**: Are the expressions coherent, consistent with the problem description, and logically valid?
   - **Clarity**: Is the LaTeX notation and overall structure easy to follow and aligned with standard optimization conventions?
3. **Rank** the solutions:
   - **"Best" (rank=1)**: The most complete, correct, and clear.
   - **"Second_Best" (rank=2)**: Good, but weaker than the best.
4. **Important**:
   - Any candidate (1 or 2) could be best.
   - Do not automatically select Candidate 1 as best.
   - Provide each rank only once (1, 2).
5. At the end, output your final ranking strictly in the JSON format we require (with keys "Best" and "Second_Best").
Problem Description:
Candidate 1:
Candidate 2:
Now, based on your thorough evaluation, provide your final ranking as valid JSON.
\end{lstlisting}

\end{document}